\documentclass[conference]{IEEEtran}

\usepackage{graphicx,amsmath,amssymb,color,multirow}
\usepackage{algorithmicx}
\usepackage{algpseudocode}
\usepackage{cases}
\usepackage{epsfig}
\usepackage{ulem}
\usepackage{epstopdf}
\usepackage{listings}
\usepackage{float}
\usepackage{makecell}
\usepackage{tabularx}
\usepackage{tikz}
\usepackage[acronym]{glossaries}
\usepackage[bookmarks=false]{hyperref}
\usepackage{setspace}
\usepackage{url}
\usepackage{breakurl}
\ifCLASSOPTIONcompsoc
\usepackage[caption=false,font=normalsize,labelfont=sf,textfont=sf]{subfig}
\else
\usepackage[caption=false,font=footnotesize]{subfig}
\fi

\graphicspath{ {Images/} }

\makeatletter
\def\ps@IEEEtitlepagestyle{%
	\def\@oddfoot{\mycopyrightnotice}%
	\def\@evenfoot{}%
}
\def\mycopyrightnotice{%
	{\begin{minipage}{2\linewidth}\footnotesize\bfseries \copyright 2020 IEEE.  Personal use of this material is permitted.  Permission from IEEE must be obtained for all other uses, in any current or future media, including reprinting/republishing this material for advertising or promotional purposes, creating new collective works, for resale or redistribution to servers or lists, or reuse of any copyrighted component of this work in other works.\hfill\end{minipage}}
	\gdef\mycopyrightnotice{}
}

\begin{document}
	 \title{Time-to-Provision Evaluation of IoT Devices Using Automated Zero-Touch Provisioning}
	
	\author{Ivan Bo\v skov$^{\ast\dagger}$,
		Halil Yetgin$^{\ast\ddagger}$,
		Matev\v z Vu\v cnik$^{\ast\dagger}$,
		Carolina Fortuna$^{\ast}$,
		Mihael Mohor\v ci\v c$^{\ast\dagger}$\\
		$^{\ast}$Department of Communication Systems, Jo{\v z}ef Stefan Institute, SI-1000 Ljubljana, Slovenia.\\
		$^{\dagger}$Jo{\v z}ef Stefan International Postgraduate School, Jamova 39, SI-1000 Ljubljana, Slovenia.\\
		$^{\ddagger}$Department of Electrical and Electronics Engineering, Bitlis Eren University, 13000 Bitlis, Turkey.\\
		\{ivan.boskov $\mid$ halil.yetgin $\mid$ matevz.vucnik $\mid$ carolina.fortuna $\mid$ miha.mohorcic\}@ijs.si    
	}
	
	\maketitle
	
	\begin{abstract}
		The Internet of Things (IoT) is being widely adopted in today's society, interconnecting smart embedded devices that are being deployed for indoor and outdoor environments, such as homes, factories and hospitals. Along with the growth in the development and implementation of these IoT devices, their simple and rapid deployment, initial configuration and out-of-the-box operational provisioning are becoming prominent challenges to be circumvented. Considering a large number of heterogeneous devices to be deployed within next generation IoT networks, the amount of time needed for manual provisioning of these IoT devices can significantly delay the deployment and manual provisioning may introduce human-induced failures and errors. By incorporating zero-touch provisioning (ZTP), multiple heterogeneous devices can be provisioned with less effort and without human intervention. In this paper, we propose software-enabled access point (Soft-AP)- and Bluetooth-based ZTP solutions relying only on a single mediator device and evaluate their performances using LOG-A-TEC testbed against manual provisioning in terms of the time required for provisioning (time-to-provision, TTP). We demonstrate that on average, Soft-AP- and Bluetooth-based ZTP solutions outperform manual provisioning with about 154\% and 313\% when compared to the expert provisioning, and with about 434\% and 880\% when compared to the non-expert provisioning in terms of TTP performances, respectively.
	\end{abstract}
	
	%
	\IEEEpeerreviewmaketitle
	
	%
	\IEEEpeerreviewmaketitle
	
	
	\section{Introduction}
	\label{sec:intro}
	The Internet of Things (IoT) has stimulated new trends and empowered innovative smart devices, which are tightly interconnecting our physical world through many indoor and outdoor applications~\cite{7123563Fuqaha}. With the growth in the development and implementation of these smart devices, their simple and rapid deployment, initial configuration and out-of-the-box operational provisioning represent prominent challenges. For example, according to a recent whitepaper from a device manufacturer, provisioning ten	thousand smart light bulbs in a factory can take nearly 2 years before they can actually commence data stream~\cite{Associates}. Therefore, to realize efficient and sustainable large scale IoT deployments, also for the fifth generation (5G) and beyond, automated provisioning such as zero-touch provisioning (ZTP)~\cite{Associates} is needed. 
	
	In its essence, ZTP is an automation solution that is designed to reduce errors and cut down the required time when network administrators need to bring new infrastructure online, thus to avoid manual provisioning process~\cite{demchenko2015enabling}. Therefore, ZTP enables accomplishing and providing remote device provisioning, where even an end-user that does not have the technical knowledge can simply connect a non-provisioned device to the network~\cite{demchenko2015enabling}. Having reviewed the paucity of the literature and the industry perspectives, the characteristics of a typical ZTP can be summarized as follows.
	
	\begin{itemize}
		\item Ease of Use - provisioning could be performed by anyone even without technical background.
		\item Interoperability - ZTP should be independent of vendors.
		\item Security - during the provisioning process, credentials should not be compromised.
		\item Ease of Implementation - provisioning mechanism should be simple to implement without requiring additional equipment and complex infrastructure.
		\item Scalability - since future generation networks are constantly evolving, provisioning process should adapt to the changes in the existing network, e.g. additional devices, capacity expansion and more sophisticated services.
	\end{itemize}
	
	Among these characteristics, our provisioning solution mainly focuses on interoperability, ease of use and ease of implementation. We propose a ZTP solution that the network devices can be provisioned with a single automated process requiring only minimal human intervention and without bringing any additional equipment, which is potentially the most favorable way to address the manual provisioning challenge. The main contributions of this paper are outlined as follows.
	
	\begin{itemize}
		\item We propose automated software-enabled access point (Soft-AP)- and Bluetooth-based ZTP solutions relying only on a single mediator device, which do not require additional hardware, while mainly considering the interoperability, ease of use and ease of implementation characteristics of a typical ZTP.
		\item Our ZTP solutions are vendor independent and can be applied to any IoT device having wireless connectivity and/or without external input/output capability, which renders the solutions to be interoperable.
		\item We evaluate the performance of the proposed automated ZTP solutions on a testbed against manual provisioning in terms of the required amount of time (time-to-provision, TTP) during which an IoT device can be provisioned.
	\end{itemize}{}
	
	The rest of this paper is organized as follows. In Section~\ref{sec:mot-back}, we provide the motivation and relevant background for ZTP. Section~\ref{sec:implementation} describes our proposed zero-touch provisioning solutions using both WiFi and Bluetooth, while experimental details and performance evaluation are presented in Section~\ref{sec:experiment}. We conclude the paper in Section~\ref{sec:conclusion}.

	\section{Motivations and Background}
	\label{sec:mot-back}

	Suppose a user purchases an IoT device and plans to deploy it within a smart environment. Perhaps, the first thing would be to configure the device to have Internet access, and secondly to manually provision the device on-site~\cite{cheshire2005dynamic}. The latter process would require a considerable amount of time if the user is not familiar with connected IoT devices. Hence, every new device integrated into the network is expected to introduce additional setup delay, which may ultimately become a tedious process. ZTP, in this regard, does not only aid in reducing the deployment delay and human-induced errors but also allows users to focus their attention on other operational tasks, which eventually improves the reliability and user experience.
	
	Traditional provisioning has been conducted by manual handling of the device prior to deployment, which often requires an expert of the system and the device to be provisioned. Additionally, as the number of connected IoT devices grows exponentially, a key obstacle in reaching envisioned 50 billion and more connected devices would be the current manual provisioning methods~\cite{evans2011internet,ericsson}. To circumvent these hindrances, we aim to provide ZTP solutions for IoT devices in an efficient and automated way with minimum human intervention, which may become particularly challenging due to small unobtrusive sensors having limited external input and output capability, and constrained physical accessibility. 
	
	The ZTP solutions for associating device to the network can be divided into two main categories, i.e., \textit{standard- and vendor-based solutions}, which are based on communication standards, such as WiFi, and are vendor-dependent, such as proprietary hardware and software, respectively.
	
	\subsection{Standard-based solutions}
	One of the commonly-used industry standards available today are WiFi protected setup (WPS)~\cite{viehbock2011brute} and Push-Button-Connect (PBC)~\cite{viehbock2011brute}, which require either an external input interface or physical access within a short distance to configure devices. To mitigate such dependencies, the most widely-used approach to configure a device is made possible via the software-enabled access point (Soft-AP) solution~\cite{iotivity}. In this solution, a non-provisioned device generates its own network making it discoverable, therefore granting a computer or a smartphone to connect to it directly and potentially facilitates its initial configuration via web browser. The smartphone or computer does not necessarily need to be connected to the Internet during the configuration procedure~\cite{iotivity}. Using the credentials provided by the user, the device can connect to the Internet. There are, however, other means to improve this mechanism by employing a QR code with the device information so that the client can avoid typing credentials~\cite{liu2008recognition}. By scanning this QR code with a mediator device, which is the device acting as the configurator, we can thereafter send the configuration file to the device. Generally speaking, these solutions often  require additional hardware components, external input and output capability and physical accessibility, which are usually infeasible for small-sized unobtrusive sensors.
	
	\subsection{Vendor-based solutions}
	Vendor- and proprietary-based solutions require applications and other means of communication to accomplish the interaction between end-user and the device. These solutions are mostly vendor-specific and are tailored to a specific range of devices. A large number of network equipment vendors, such as Cisco, Juniper, Apple and Texas Instruments, partly implement ZTP services into their devices. For example, Cisco's ZTP feature provides a solution where a non-provisioned device communicates to a configuration engine in order to automatically retrieve a complete configuration file~\cite{ciscoztp}. Juniper Networks provides a device-specific provisioning based on the location of the devices, which is achieved by leveraging Juniper Networks built-in functions~\cite{juniperztp}. Apple developed a software tool that allows the client to utilize their Apple products as a mediator device allowing to broadcast the configuration to nearby devices via its proprietary solution to seamlessly connect its smart devices to a network by using Apple wireless accessory configuration (WAC) feature~\cite{AppleWAC}. Texas Instruments developed the SmartConfig technology~\cite{reiter2014primer}, which utilizes a mobile or computer application to write the password of the network on which the mediator device is connected to. However, one significant drawback of this solution is that it relies on an application that needs to be preinstalled on a smartphone or a computer. Generally speaking, most of these solutions are strictly dependent on the proprietary software or additional prebuilt dependencies, protocol and hardware of specific vendors.
	

	\begin{figure*}[htb]
		\includegraphics[scale=0.5]{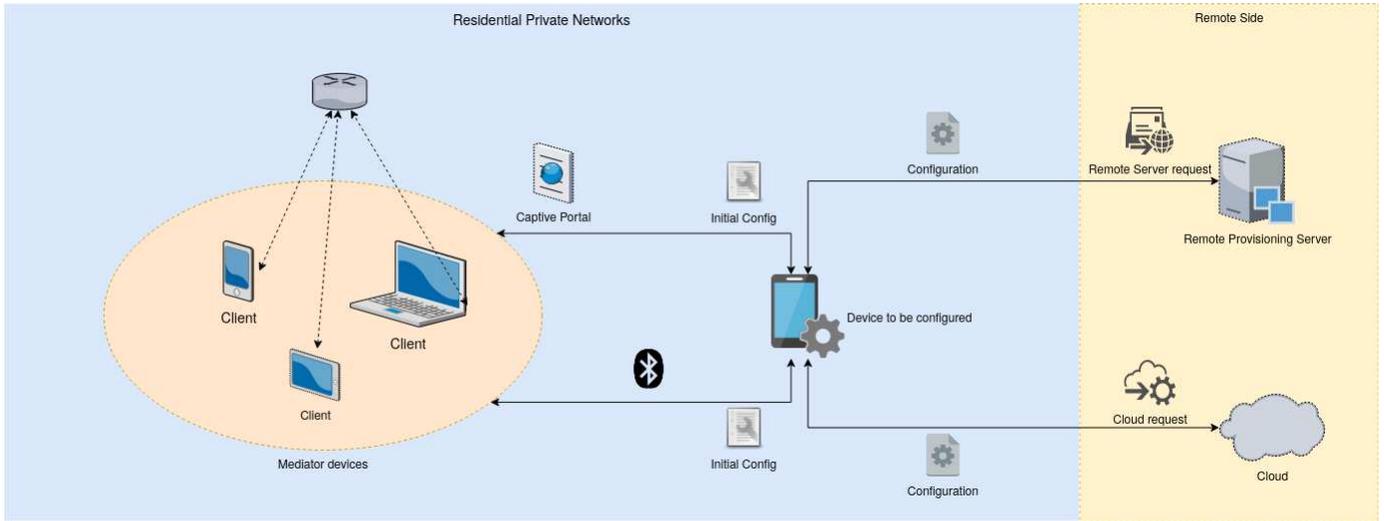}
		\centering
		\caption{Architecture of the customized Soft-AP- and Bluetooth-based ZTP solutions portraying a remote configuration file access.}
		\label{fig:systemarch}
	\end{figure*}
	
	\section{Proposed ZTP Solutions}
	\label{sec:implementation}
	
	\begin{figure}[!thb]
		\includegraphics[width=\columnwidth]{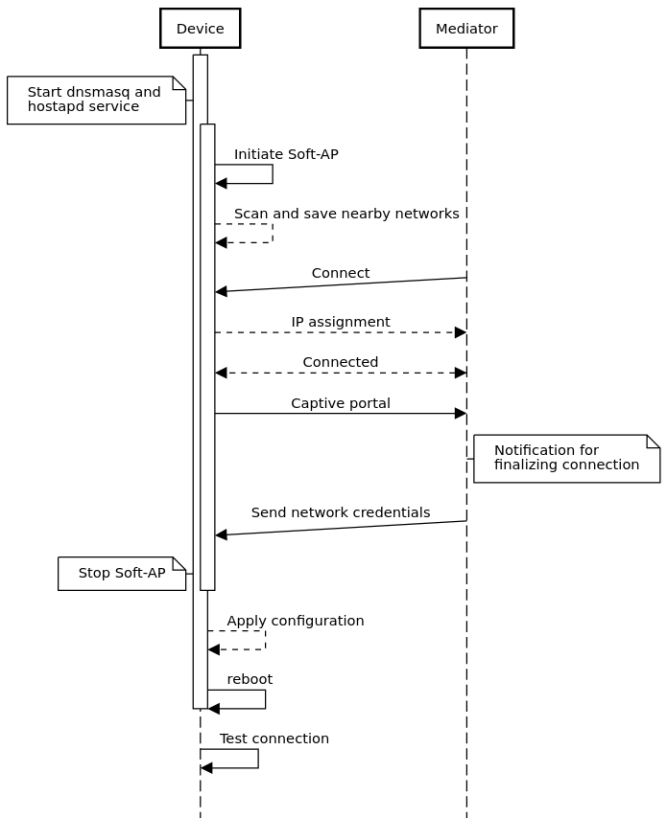}
		\centering
		\caption{Sequence flowchart of the customized Soft-AP-based ZTP solution.}
		\label{fig:softap}
	\end{figure}
	The aforementioned provisioning solutions are all targeted towards a specific type of device and fail to accomplish a general solution for heterogeneous IoT devices. Under this premise, the goal of this paper is to propose innovative ZTP solutions based on WiFi and Bluetooth technologies, which can yield sufficient performance for provisioning of multiple devices without depending on the vendor's proprietary hardware and software, while fostering a relatively simple deployment and enabling an efficient provisioning process. 
	
	The proposed solution includes a single mediator device that is also not strictly enforced to have an Internet access, essentially allowing for a non-provisioned device to register with the network and fetch the up-to-date configuration file. In our developed software tool\footnote{All scripts are available at: https://github.com/iboskov/ZeroFi-snippets.} for the proposed ZTP solutions, we opted for adopting the Soft-AP method~\cite{iotivity} and attained a better automated ZTP solution via Bluetooth. Moreover, the proposed provisioning tool is software-based and does not rely on specific hardware.
	
	Fig.~\ref{fig:systemarch} depicts the architecture of the proposed solutions on how the provisioning is carried out. The mediator device can be a laptop, a tablet or a smart phone, and they do not require prior Internet connection, as portrayed via dashed lines of Fig.~\ref{fig:systemarch}. Additionally, the two approaches to initialize the provisioning, whether through WiFi or Bluetooth, are provided in Fig.~\ref{fig:systemarch}, which also portrays the remote server and cloud requests for a configuration file upon provisioning by using one of our proposed solutions. In all scenarios, we assume that the devices are pre-installed with a basic operating system.
		

\subsection{Customized Soft-AP provisioning}
In Soft-AP approach, when the device boots up, initially it sets up the \textit{dnsmasq} and \textit{hostapd} services, which are responsible for redirecting the user to a configuration web page and enabling the Soft-AP, respectively. The solution in~\cite{iotivity} utilizes Soft-AP to allow devices to be discoverable, therefore mediators can connect to the network of these respective devices. However, this procedure requires an application to carry out the configuration. Fig.~\ref{fig:softap} depicts the flow chart of the provisioning of the device with the aid of Soft-AP, which is automated and occurs in the background. Therefore, the entire process is invisible to the person carrying out the configuration. Furthermore, booting into Soft-AP mode will allow a user to directly connect to the device with the help of a mediator device. 
	
Upon establishing an initial connection, the user will receive a notification on his/her device, which will redirect them to a configuration web page when opened. This is achieved with the help of \textit{Captive Portal}, which represents a web page accessed through a web browser that is displayed to newly connected users of a WiFi network before they are granted broader access. These captive portals can be utilized to provide access to enterprise or residential wired networks, such as apartments, hotel rooms and business centers. In our case, the device includes an embedded web server on which the captive portal is hosted. The web-based form either automatically opens in a web browser or appears when the user opens a web browser and attempts to visit a web page. This indicates that the user is “captive” and unable to access the Internet freely until the required form at the captive portal is completed. 
	
We leverage and improve upon this regular Soft-AP approach by means of customizing the captive portal so as to display a list of all nearby WiFi networks around the device, unlike~\cite{iotivity}, where the device is only discoverable through Soft-AP and does not utilize Captive Portal to be configured, nor scans for nearby available networks. Besides, there is an option to enter the network manually if the network is set to be hidden. Normally in regular soft-AP approach, if the credentials are incorrect, the device may be blocked and may require a factory reset. However, in our customized Soft-AP solution, all of these procedures are automated, where, for example, in case of an incorrect passphrase, it will revert back and wait for the correct credentials. Furthermore, selecting the desired network and typing the correct password will set off a flag implying that the device will reboot as a client, which then initiates tests to verify that an active Internet connection is established. 
	
During the proposed ZTP process, the end-user does not need to have an Internet access on their mediator device, granting for a complete offline setup and registering the device to a local network. At the end of this procedure, we have effectively registered a non-provisioned device to a local network allowing it to perform requests for ZTP. However, Soft-AP provisioning solution still requires a user to manually enter credentials via captive portal. To circumvent this drawback, we propose a fully-automated Bluetooth-based ZTP solution described as follows.
	
	\begin{figure}[!htb]
		\includegraphics[width=\columnwidth]{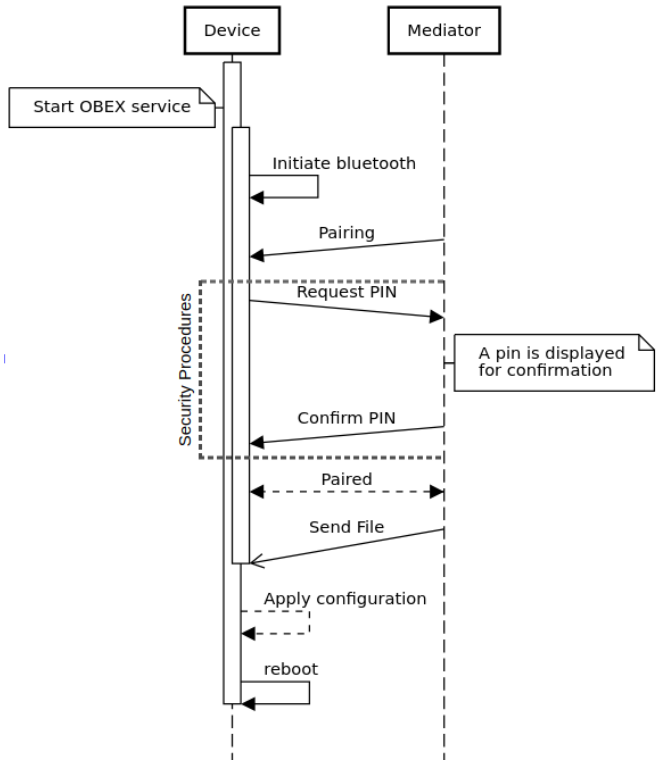}
		\centering
		\caption{Sequence flowchart of the Bluetooth-based ZTP solutions including (Secure-Bluetooth) and excluding security procedures depicted by dotted lines.}
		\label{fig:bluetooth}
	\end{figure}

	\subsection {Bluetooth provisioning}
	In this proposed solution, we make use of the readily available Bluetooth technology for ZTP, where the flowchart of our implementation and provisioning process are illustrated in Fig.~\ref{fig:bluetooth}. This provisioning process is relatively simpler compared to our customized Soft-AP solution, since we do not need any web server to enter manual credentials nor an initial configuration. Instead, once the device boots up, it is ready to pair. This configuration works with a Bluetooth agent specifying the security method\footnote{Bluetooth scenario excluding \textit{security procedures}, illustrated in Fig.~\ref{fig:bluetooth} with dotted lines, is also available through the scripts at: https://github.com/iboskov/ZeroFi-snippets.}, whether it is a PIN or a passphrase. However, for the sake of simplicity, we have opted for using a PIN which will be displayed on the user's device for verification. Furthermore, once the initial pairing is accomplished, the device to be configured and the mediator device are enabled for transferring files, such as sending network credentials to initiate ZTP or transferring configuration files that would append distinct operational capabilities to the device. 
	
	Upon the boot up, the device runs object exchange (OBEX) protocol as a service that enables transfer of data between two devices over Bluetooth, as depicted in Fig.~\ref{fig:bluetooth}. However, this would be inapplicable if the Bluetooth of the device to be configured is not initialized. Upon initialization of the Bluetooth, the device can accept and approve the incoming connections. Following a successful connection, the device initially verifies if the connecting device is in the list of trusted devices. If the device has already been paired previously, it is expected to automatically connect. However, if this is the first connection to be established, then the pairing process will begin before a seamless connection takes place.
	
	When the secure connection between the two devices has been established upon realizing the ZTP, the device to be configured awaits a configuration file from the mediator device. This file can be highly complex and for the time being, a basic text file parser and an HTML parser are already implemented in our developed software tool so that the device is able to interpret the configuration scripts. After receiving the configuration file, the device will confirm the format of the file, whether it is a text file or an HTML string, so that it can proceed to determine the adequate means for implementing the changes. At the end of this procedure, the device tests whether the configuration was successful.

	\section{Performance Analysis}
	\label{sec:experiment}
	To evaluate the performance of the proposed solutions, in this section we first provide details on the experimental setup, then we elaborate on the methodology used for evaluation and finally present and discuss the results.
	  
	\subsection{Experimental details}
\begin{figure}[!htb]
		\includegraphics[width=0.932\columnwidth]{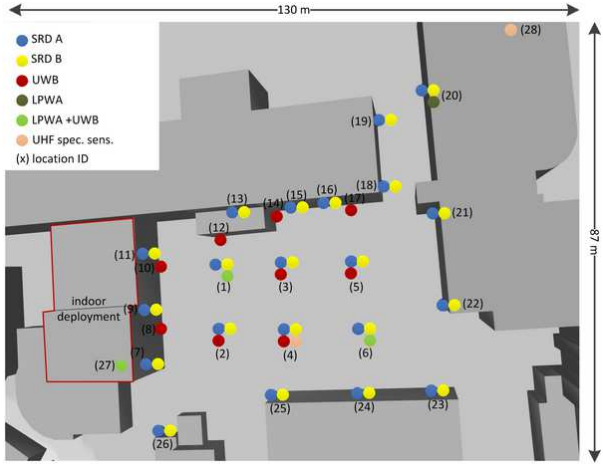}
		\centering
		\caption{The structural map of the testbed.}
		\label{fig:testbed}
	\end{figure}

	To realistically evaluate the performance, the customized Soft-AP- and Bluetooth-based ZTP solutions are implemented over part of the LOG-A-TEC\footnote{http://www.log-a-tec.eu} testbed at Jo\v zef Stefan Institute, as illustrated in Fig.~\ref{fig:testbed}. The outdoor part of the wireless testbed is located in and around a park area of 55 m by 60 m. The testbed is extended to support both indoor and outdoor scenarios with additional 20 ultra-wideband (UWB) devices and one low-power wide-area (LPWA) device mounted at the second and third floors of a 28.4 m by 16.6 m building.
	
\begin{figure}[!htb]
	\includegraphics[width=0.932\columnwidth]{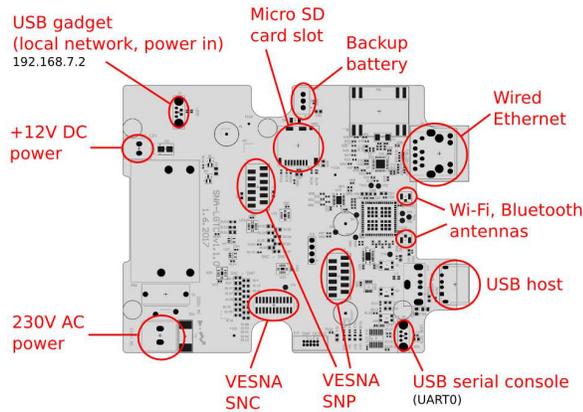}
	\centering
	\caption{Layout of SNA-LGTC board.}
	\label{fig:testbeddev}
\end{figure}

	The devices used for testing provisioning solutions are custom made SNA-LGTC boards\footnote{https://github.com/sensorlab/sna-lgtc-support}. This is a small, ARM-based Linux-running computer that is capable of hosting microcontroller-based boards. The board has a single core 1GHz ARMv7 processor, 512MB RAM memory with 4GB 8-bit eMMC on-board flash storage and wireless capabilities IEEE 802.11abgn, and integrated with Bluetooth v4.1. The layout of the board along with its connectors is portrayed in Fig.~\ref{fig:testbeddev}. Since the device supports both Bluetooth and WiFi, it will be used for evaluating both provisioning solutions.
	
	\subsection{Evaluation methodology}
\begin{figure}
		\includegraphics[width=\columnwidth]{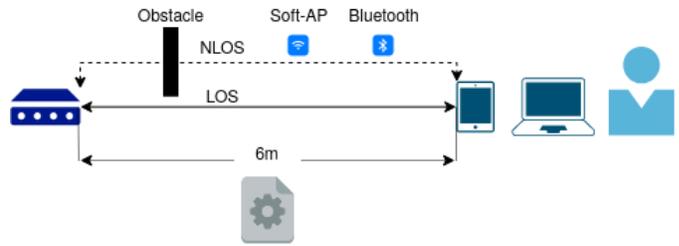}
		\caption{Per-link time-to-provision (TTP) evaluations for both the customized Soft-AP- and Bluetooth-based provisioning solutions.}
		\label{fig:softap-blue-link}
	\end{figure}

	As the amount of time required for provisioning (time-to-provision, TTP) is the most important key metric in this context, we leverage TTP to evaluate the performance of the customized Soft-AP- and Bluetooth-based provisioning solutions. To fairly evaluate and compare the performance of the customized Soft-AP- and Bluetooth-based provisioning solutions, we opt for the evaluation of a single link (per-link) of a short-range device (SRD A) acquired from Fig.~\ref{fig:testbed} for both line-of-sight (LOS) and non-LOS (NLOS) scenarios, as appropriately arranged in the form of Fig.~\ref{fig:softap-blue-link}, which also presents the distance of 6 [m] set between mediator and the device to be configured. For the evaluation of customized Soft-AP ZTP solution, a laptop with 5 GHz WiFi band is utilized as a mediator device. In case of the evaluation of Bluetooth-based ZTP solution, smartphones with various Bluetooth versions are leveraged as mediator devices, such as Samsung Galaxy Note 10+ - Bluetooth 5.0 and LG Nexus 4 - Bluetooth v4.0, which are then evaluated for the sake of interoperability. By default, Bluetooth v5.0 without security procedures (non-secure) is utilized for the experimental evaluation, unless stated otherwise. We evaluate the TTP performance for the following scenarios, each of which is averaged for 15 evaluation tests when automated provisioning solutions are considered. Additionally, manual provisioning scenario is divided into two cases; i) provisioning is conducted by one expert over 15 evaluations, and ii) provisioning is conducted by 15 different non-experts and each executed one evaluation test. The latter case is preferred since a non-expert can learn the provisioning guidelines and he/she may become an expert till the end of the evaluation tests. Therefore, we took this step as a countermeasure for maintaining the fairness of the evaluation.
	
\begin{table*}[!htb]
	\renewcommand{\arraystretch}{1.2}
	\centering
	\caption{TTP evaluation results averaged over 15 tests for the automated ZTP solutions as well as for the manual expert provisioning case, whereas manual non-expert provisioning case accounts for 15 different non-experts each executing one evaluation test.}
	\label{tab:eval}
	\begin{tabular}{|l|c|c|c|c|c|c|c|c|} 
		\hline
		\multirow{2}{*}{Evaluation Scenarios} & \multicolumn{2}{c|}{Manual Provisioning} & \multirow{2}{*}{LOS-AP} & \multirow{2}{*}{NLOS-AP} & \multirow{2}{*}{LOS-BL} & \multirow{2}{*}{NLOS-BL} & \multirow{2}{*}{LOS-S-BL} & \multirow{2}{*}{LOS-BL4} \\ 
		\cline{2-3}
		& Expert & Non-expert &  &  &  &  &  &  \\ 
		\hline
		\hline
		Average TTP [sec] & 46.88 & 131.87 & 30.38 & 43.87 & 14.98 & 28.34 & 25.54 & 15.63 \\ 
		\hline
		Best-effort TTP [sec] & 44.83 & 126.31 & 27.57 & 40.79 & 10.33 & 24.94 & 21.58 & 11.36 \\ 
		\hline
		Worst-effort TTP [sec] & 50.71 & 193.23 & 39.32 & 48.80 & 17.97 & 32.70 & 29.88 & 17.91 \\
		\hline
	\end{tabular}
\end{table*}

\begin{figure*}[tbp]
	\centering
	\includegraphics[width=1.01\linewidth,height=5.4cm]{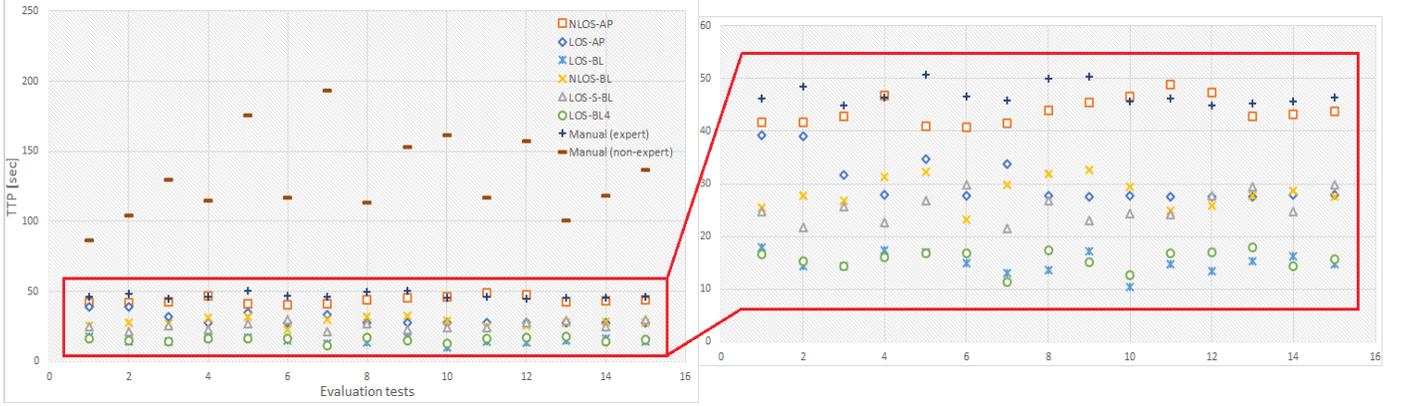}
	\caption{Distribution of TTP evaluation tests for all provisioning scenarios.}
	\label{fig:dist}
\end{figure*}
	
\begin{figure}
	\centering
	\includegraphics[width=0.95\columnwidth]{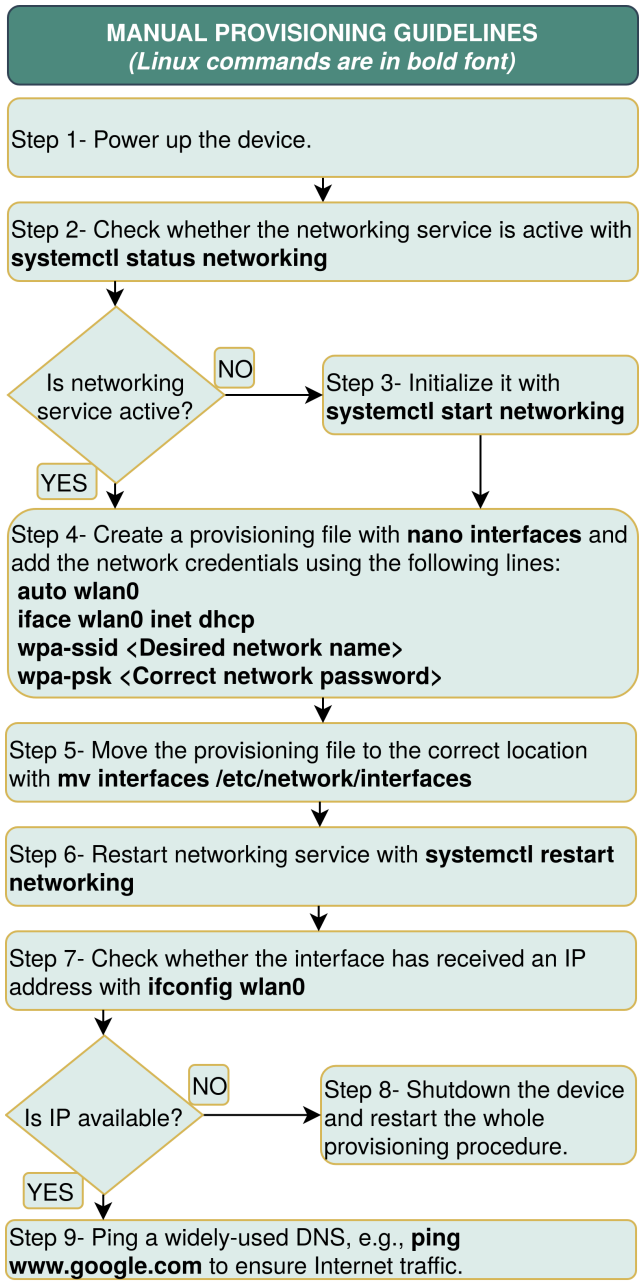}
	\caption{Step-by-step manual provisioning guidelines.}
	\label{fig:manprov}
\end{figure}

	\begin{enumerate}
		\item \textit{Manual Provisioning (Manual)} is a baseline evaluation method to explicitly understand the performance improvements that the automated ZTP solutions put forward, where the device will be provisioned by one expert over 15 times who is familiar with the provisioning procedures, and by 15 other non-experts with no previous knowledge following a step-by-step provisioning guideline that is provided in Fig.~\ref{fig:manprov}. Note that along with the guidelines, we made a few introductory remarks for each non-expert before commencing with the manual provisioning.
		\item \textit{LOS-Soft-AP (LOS-AP)} is to evaluate the customized Soft-AP-based ZTP solution for LOS scenario using WiFi.
		\item \textit{NLOS-Soft-AP (NLOS-AP)} is to evaluate the customized Soft-AP-based ZTP solution for NLOS scenario using WiFi.
		\item \textit{LOS-Bluetooth (LOS-BL)} is to evaluate the Bluetooth-based (V5.0) ZTP solution for LOS scenario without security procedures considered.
		\item \textit{NLOS-Bluetooth (NLOS-BL)} is to evaluate the Bluetooth-based (V5.0) ZTP solution for NLOS scenario without security procedures considered.
		\item \textit{LOS-Secure-Bluetooth (LOS-S-BL)} is to evaluate the Bluetooth-based (V5.0) ZTP solution for LOS scenario including security procedures.
		\item \textit{LOS-Bluetooth-v4.0 (LOS-BL4)} is to evaluate the Bluetooth-based (V4.0) ZTP solution for LOS scenario without security procedures considered for exemplifying the interoperability.
	\end{enumerate}

\subsection{Evaluation results and discussions}
	
The distribution of the TTP evaluation tests are scattered on Fig.~\ref{fig:dist}, and average, best-effort (to potentially reveal how much time is required for manual provisioning conducted by an expert with advanced knowledge when compared to the automated ZTP solutions) as well as worst-effort (to potentially reveal how much time is required for manual provisioning performed by a non-expert without technical background when compared to the expert counterpart case and to the automated ZTP solutions) TTP evaluation performances are provided in Table~\ref{tab:eval}. Evaluation results of Table~\ref{tab:eval} reveal that even the best-effort TTP (44.83 secs) of manual provisioning solely performs better with about 9\% margin than the worst-effort TPP (48.80 secs) among all the automated provisioning scenarios and on average, Soft-AP-based (LOS-AP, 30.38 secs) and Bluetooth-based (LOS-BL, 14.98 secs) solutions outperform manual provisioning for the cases of an expert and non-experts (46.88 and 131.87 secs) with about 154\% and 313\% for the expert provisioning, and with about 434\% and 880\% for the non-expert provisioning in terms of TTP performances, respectively. This reveals that the automated ZTP solutions outperform the best manual TTP performance with about at least 154\%. 
	
For the automated ZTP solutions, on average, LOS introduces nearly 44\% of TTP improvement for the Soft-AP-based ZTP solutions and about 89\% improvement for the Bluetooth-based ZTP solutions, when compared to their respective NLOS counterparts. Moreover, taking security procedures into account for Bluetooth-based ZTP solutions, namely LOS-S-BL, an average of 42\% TTP performance degradation is recorded compared to non-secure Bluetooth-based ZTP solutions. Nonetheless, a device having Bluetooth version v4 introduced a tiny margin of 4\% TTP performance degradation when compared to Bluetooth v5, which can potentially present the interoperability of our Bluetooth-based ZTP solutions. 
	
Generally speaking, all the proposed Bluetooth-based ZTP solutions outperformed the Soft-AP-based ZTP solutions, particularly considering the best scenarios of both, the LOS-BL presents about 103\% better TTP performance than the LOS-AP. These discussed trends can also be readily observed in Fig.~\ref{fig:dist} demonstrating all the evaluation tests, which also proves the convergence of the automated ZTP solutions. Note that even with the worst-effort TTP performance (NLOS-AP, 48.80 secs), the user has the flexibility to focus their attention on other operational tasks, while this is impractical for manual provisioning.

\section{Conclusions}
\label{sec:conclusion}
In this paper, we proposed Soft-AP- and Bluetooth-based ZTP solutions improved upon the well-known Soft-AP method, which aim for reducing the required TTP and human-induced failures by automatizing the entire procedure of manual provisioning. To this end, our proposed ZTP solutions along with the developed lightweight software tool can provision IoT devices at the expense of very few resources (only requiring wireless connectivity and a single mediator device) approving the ease of use, ease of implementation and interoperability (no vendor dependency) characteristics of the ZTP. The proposed ZTP solutions are evaluated on a per-link basis considered as part of the LOG-A-TEC testbed in terms of TTP performance, which is perhaps the most important evaluation metric for such scenarios. We proved the effectiveness of our proposed ZTP solutions over manual provisioning and concluded that Bluetooth-based ZTP solutions outperformed all the other ZTP solutions proposed. As a future work, we plan to evaluate the scalability of our ZTP solutions by means of implementing over the entire testbed.
	
\section*{Acknowledgment}
This work was funded by the Slovenian Research Agency (Grant no. P2-0016).
	\bibliography{bibliography}

\begin{thebibliography}{10}
\providecommand{\url}[1]{#1}
\csname url@samestyle\endcsname
\providecommand{\newblock}{\relax}
\providecommand{\bibinfo}[2]{#2}
\providecommand{\BIBentrySTDinterwordspacing}{\spaceskip=0pt\relax}
\providecommand{\BIBentryALTinterwordstretchfactor}{4}
\providecommand{\BIBentryALTinterwordspacing}{\spaceskip=\fontdimen2\font plus
\BIBentryALTinterwordstretchfactor\fontdimen3\font minus
  \fontdimen4\font\relax}
\providecommand{\BIBforeignlanguage}[2]{{%
\expandafter\ifx\csname l@#1\endcsname\relax
\typeout{** WARNING: IEEEtran.bst: No hyphenation pattern has been}%
\typeout{** loaded for the language `#1'. Using the pattern for}%
\typeout{** the default language instead.}%
\else
\language=\csname l@#1\endcsname
\fi
#2}}
\providecommand{\BIBdecl}{\relax}
\BIBdecl

\bibitem{7123563Fuqaha}
A.~{Al-Fuqaha}, M.~{Guizani}, M.~{Mohammadi}, M.~{Aledhari}, and M.~{Ayyash},
  ``Internet of things: A survey on enabling technologies, protocols, and
  applications,'' \emph{IEEE Communications Surveys and Tutorials}, vol.~17,
  no.~4, pp. 2347--2376, Forthquarter 2015.

\bibitem{Associates}
\BIBentryALTinterwordspacing
J.~Wilhelm, J.~Williams, and S.~Macy. (2017) Whitepaper on {IoT} onboarding - a
  device manufacturer’s perspective: Addressing a keycustomer need to onboard
  {IoT} devices securely and efficiently. [Online]. Available:
  \url{https://www.intel.com/content/dam/www/public/us/en/documents/white-papers/kaiser-associates-iot-onboarding-for-device-manufacturers-whitepaper.pdf}
\BIBentrySTDinterwordspacing

\bibitem{demchenko2015enabling}
Y.~Demchenko, S.~Filiposka, R.~Tuminauskas, A.~Mishev, K.~Baumann, D.~Regvart,
  and T.~Breach, ``Enabling automated network services provisioning for cloud
  based applications using zero touch provisioning,'' in \emph{IEEE/ACM 8th
  International Conference on Utility and Cloud Computing (UCC)}, Limassol,
  Cyprus, December 2015.

\bibitem{cheshire2005dynamic}
S.~Cheshire, B.~Aboba, and E.~Guttman, ``Dynamic configuration of {IPv4}
  link-local addresses,'' RFC 3927 (Proposed Standard), Tech. Rep., 2005.

\bibitem{evans2011internet}
D.~Evans, ``The internet of things: How the next evolution of the internet is
  changing everything,'' \emph{CISCO white paper}, 2011.

\bibitem{ericsson}
\BIBentryALTinterwordspacing
Ceo to shareholders: 50 billion connections 2020. (accessed: 28.10.2019).
  [Online]. Available:
  \url{https://www.ericsson.com/en/press-releases/2010/4/ceo-to-shareholders-50-billion-connections-2020}
\BIBentrySTDinterwordspacing

\bibitem{viehbock2011brute}
S.~Viehb{\"o}ck, ``Brute forcing wi-fi protected setup,'' \emph{Wi-Fi Protected
  Setup}, vol.~9, 2011.

\bibitem{iotivity}
\BIBentryALTinterwordspacing
Iotivity easy setup, wifi provisioning. (accessed: 28.10.2019). [Online].
  Available: \url{https://wiki.iotivity.org/easy_setup}
\BIBentrySTDinterwordspacing

\bibitem{liu2008recognition}
Y.~Liu, J.~Yang, and M.~Liu, ``Recognition of {QR} code with mobile phones,''
  in \emph{IEEE Chinese control and decision conference}, Yantai, Shandong,
  China, July 2008.

\bibitem{ciscoztp}
\BIBentryALTinterwordspacing
Cisco zero-touch provisioning configuration guide, {Cisco IOS XE} 16.x.x.
  (accessed: 09.12.2019). [Online]. Available:
  \url{https://www.cisco.com/c/en/us/td/docs/switches/lan/catalyst3850/software/release/16-5/configuration_guide/prog/b_165_prog_3850_cg/zero_touch_provisioning.pdf}
\BIBentrySTDinterwordspacing

\bibitem{juniperztp}
\BIBentryALTinterwordspacing
Juniper networks configuring zero touch provisioning in branch networks.
  (accessed: 09.12.2019). [Online]. Available:
  \url{https://www.juniper.net/documentation/en_US/release-independent/nce/information-products/pathway-pages/nce/nce-151-zero-touch-provisioning.pdf}
\BIBentrySTDinterwordspacing

\bibitem{AppleWAC}
Telit, ``Gs2k wac provisioning adk application note - telit technical
  documentation - 80560nt11590a rev. 1.0,'' 2015.

\bibitem{reiter2014primer}
G.~Reiter, ``A primer to wi-fi{\textregistered} provisioning for iot
  applications,'' \emph{Texas Instruments White Paper}, 2014.

\end{thebibliography}
	\bibliographystyle{IEEEtran}

\end{document}